\documentclass[sigconf]{acmart}
\renewcommand\footnotetextcopyrightpermission[1]{} 

\usepackage{booktabs} 
\usepackage{tabularx} 
\usepackage{amsmath}
\usepackage{amssymb}
\usepackage{algorithm}
\usepackage{multirow}
\usepackage{graphicx}
\usepackage{subcaption}
\usepackage{epstopdf}
\usepackage[noend]{algpseudocode}
\usepackage{balance}

\usepackage{xspace}

\makeatletter
\def\BState{\State\hskip-\ALG@thistlm}
\makeatother

\newcolumntype{L}[1]{>{\raggedright\let\newline\\\arraybackslash\hspace{0pt}}m{#1}}
\newcolumntype{C}[1]{>{\centering\let\newline\\\arraybackslash\hspace{0pt}}m{#1}}
\newcolumntype{R}[1]{>{\raggedleft\let\newline\\\arraybackslash\hspace{0pt}}m{#1}}

\usepackage[moderate]{savetrees}

\usepackage{pifont}
\setcopyright{none}

\newif\ifmakeshorter
\makeshortertrue

\begin{document}




\title{Towards Evaluating User Profiling Methods Based on Explicit Ratings on Item Features}
\titlenote{Permission to make digital or hard copies of part or all of this work for personal or classroom use is granted without fee provided that copies are not made or distributed for profit or commercial advantage and that copies bear this notice and the full citation on the first page. Copyrights for third-party components of this work must be honored. For all other uses, contact the owner/author(s). \\ IntRS workshop, The 13th ACM Conference on Recommender Systems (RecSys), 2019,  Copenhagen, Denmark.}
 
\author{Luca Luciano Costanzo}
\affiliation{%
  \institution{Politecnico di Milano, Italy}
}
\email{lucaluciano.costanzo@mail.polimi.it}

\author{Yashar Deldjoo}
\affiliation{%
  \institution{Polytechnic University of Bari, Italy}
}
\email{deldjooy@acm.org}

\author{Maurizio Ferrari Dacrema}
\affiliation{%
  \institution{Politecnico di Milano, Italy}
}
\email{maurizio.ferrari@polimi.it}

\author{Markus Schedl}
\affiliation{%
  \institution{Johannes Kepler University Linz, Austria}
}
\email{markus.schedl@jku.at}

\author{Paolo Cremonesi}
\affiliation{%
  \institution{Politecnico di Milano, Italy}
}
\email{paolo.cremonesi@polimi.it}

\begin{abstract}
In order to improve the accuracy of recommendations, many recommender systems nowadays use side information beyond the user rating matrix, such as item content. These systems build user profiles as estimates of users' interest on content (e.g., movie genre, director or cast) and then evaluate the performance of the recommender system as a whole e.g., by their ability to recommend relevant and novel items to the target user. The user profile modelling stage, which is a key stage in content-driven RS is barely properly evaluated due to the lack of publicly available datasets that contain user preferences on content features of items. 

To raise awareness of this fact, we investigate differences between \emph{explicit} user preferences and \emph{implicit} user profiles. We create a dataset of explicit preferences towards content features of movies, which we release publicly.  
We then compare the collected {explicit} user feature preferences and {implicit} user profiles built via state-of-the-art user profiling models.
Our results show a maximum average pairwise cosine similarity of 58.07\% between the explicit feature preferences and the implicit user profiles modelled by the best investigated profiling method and considering movies' genres only. For actors and directors, this maximum similarity is only 9.13\% and 17.24\%, respectively.
This low similarity between explicit and implicit preference models encourages a more in-depth study to investigate and improve this important user profile modelling step, which will eventually translate into better recommendations. 
\end{abstract}

\ifmakeshorter
\else
\fi





\maketitle

\section{Introduction}
\label{SEC:Introduction}


The performance of collaborative filtering (CF) recommendation models have reached a remarkable level of maturity. These models are now widely adopted in real-world recommendation engines because of their state-of-the-art recommendation quality. In recent years, a number of recommendation scenarios have emerged, which have encouraged the research community to consider using various additional information sources (aka side information) beyond the user rating matrix~\cite{shi2014collaborative}. A prominent example---and the one we focus on---is item content.
In the movie domain, for instance, a variety of content features have been considered, such as metadata or features extracted directly from the core audio-visual signals. Metadata-based movie recommender systems typically use genre~\cite{DBLP:conf/ijcnn/FilhoWB17,DBLP:journals/mta/HwangPHK16,DBLP:conf/worldcist/SoaresV17} 
or user-generated tags~\cite{DBLP:conf/icmlc/LiuFY17,wei2016hybrid,zhao2017social} 
over which user profiles are built, assuming that these aspects represent the semantic content of movies. 
In contrast, audio-visual signals represent the low-level content (e.g., color, lighting, spoken dialogues, music, etc.)~\cite{DBLP:conf/recsys/DeldjooCEISC18,DBLP:journals/umuai/DeldjooDCECSIC19,DBLP:conf/cbmi/DeldjooCSQ17,DBLP:conf/ijcnn/FilhoWB17,DBLP:journals/sp/DengRQHQ18}. 
Some approaches try to infer semantic concepts from low-level representations, e.g., via word2vec embeddings~\cite{DBLP:journals/jocs/AnwaarIAN18}, deep neural networks~\cite{DBLP:journals/csur/ZhangYST19,DBLP:journals/eswa/WeiHCZT17}, fuzzy logic~\cite{DBLP:journals/jifs/VashisthKB17}, or genetic algorithms~\cite{DBLP:conf/iiwas/Mueller17}. For these reasons, it is evident that item content plays a key role in building hybrid or content-based filtering (CBF) models and, furthermore, it is important to correctly distinguish and weight the item features by their estimated relevance for a target user, to better model his or her tastes.

\begin{figure*}[!t]
    \centering
    \includegraphics[width=0.7\textwidth]{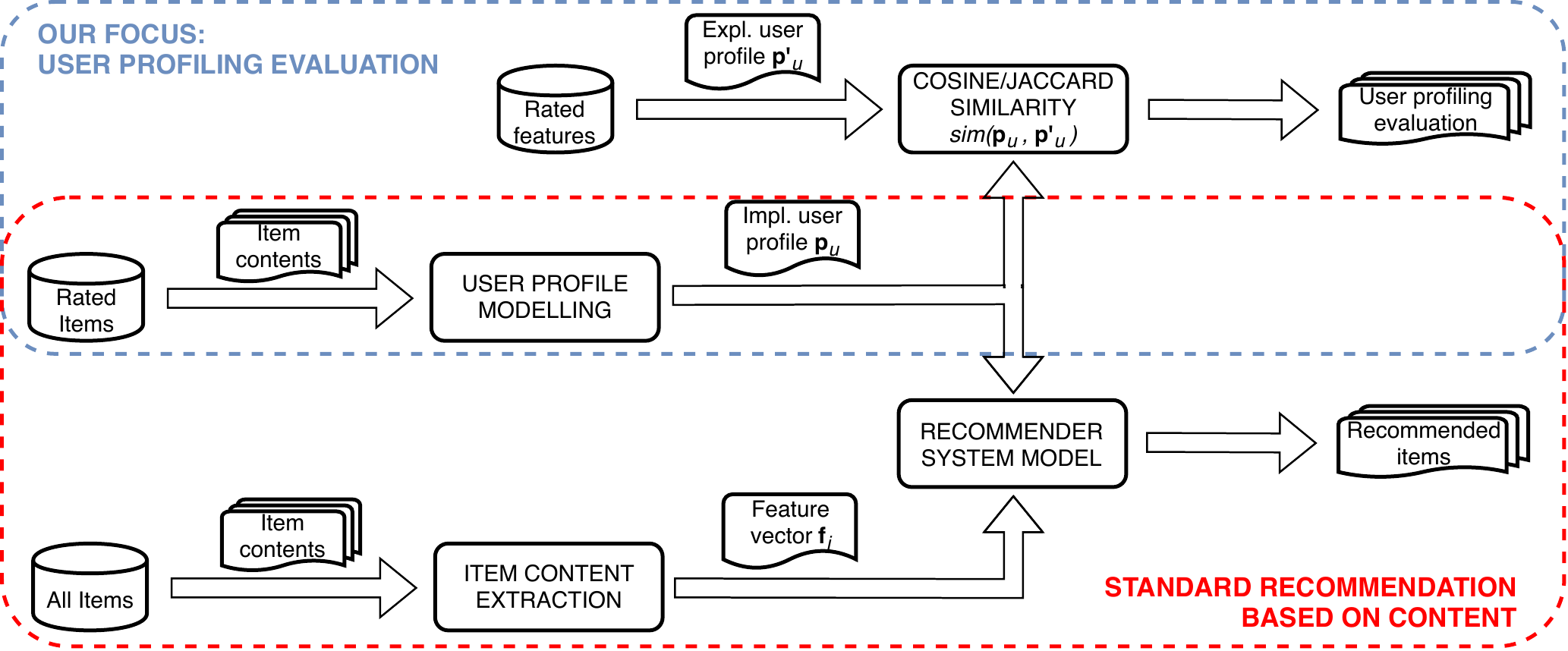}
    \caption{Main steps involved in a recommendation system leveraging content information, 
    highlighting our contributions.}
    \label{fig:standardContentRecommendation}
\end{figure*}

In Figure~\ref{fig:standardContentRecommendation}, we illustrate a simplified diagram that shows our research contributions. Standard recommendation based on content (CBF or hybrid) is structured in three main steps: (i) extraction of item content, consisting of building a \textit{feature vector}   that describes each item $i$; (ii) building the \textit{profile of the target user} $\mathbf{p}_u$, i.e., a structured representation of the user's preference over item content features;  
(iii) matching the user profile $\mathbf{p}_u$ against the feature vector of each item $\mathbf{f}_i$ to produce the list of recommended items most similar to the target user's tastes. 

A shortcoming of typical RS evaluation is that the user profiling stage, which is a key part of the RS, is barely 
evaluated. Usually, only the performance of the entire RS, which is composed of several components, is assessed and how effectively the user profiling step functions remains an open question. 
We argue that 
it is important to investigate the user profiling stage and compare performance of different profile modelling methods (see upper part of Figure~\ref{fig:standardContentRecommendation}).



The goal of this work is therefore to investigate the difference between explicit user ratings on individual movie content features (e.g., genre, actors, or directors) and implicit models inferred via state-of-the-art user modelling techniques from explicit ratings of the whole movies. To this end, we (i) create (and make publicly available) a varied dataset of explicit ratings both on movies and content features and (ii) evaluate different user profiling methods and compare their resulting implicit models against the true feature ratings provided in the collected dataset.


\section{Related Work} \label{SEC:RelatedWork}
With respect to previous research, to the best of our knowledge, the only work that evaluates implicit user profiles against true ratings on content features is~\cite{Nasery2015PoliMovie:Systems}.
Nasery et al.~compare actually rated features with the ones implicitly derived from rated movies, but no concrete user profiling methods are investigated. Instead, 
the number of times each feature is explicitly rated and the number of times it appears in the content of all rated movies is counted, and these counts are compared. 
The authors create a dataset of movies' feature ratings (genres, actors/cast, and directors), dubbed PoliMovie,\footnote{PoliMovie: \url{http://bit.ly/polimovie}} 
through a survey web application they built. Their approach, using limited survey questions and a fixed reduced dataset of top popular movies and features, extracted from IMDb,\footnote{Internet Movie Database (IMDB): \url{www.imdb.com}} tends to push users to limited and convergent preferences.
In contrast, we systematically investigate 4 methods to model implicit user profiles and we compare them with explicit user profiles obtained by feature ratings. Another contribution of the work at hand is the creation of a dataset that includes ratings on movie content features. Other datasets commonly used in movie recommender systems research, but which do not contain such feature ratings, include MovieLens 20M (ML-20M)~\cite{DBLP:journals/tiis/HarperK16}, IMDB Movies Dataset~\cite{Leka2016IMDBDataset}, The Movies Dataset~\cite{Banik2017TheMoviesDataset}, MMTF-14K and MVCD-7K~\cite{DBLP:conf/mmsys/DeldjooCISC18,DBLP:conf/cbmi/DeldjooCSQ19} and the Netflix Prize dataset~\cite{NeftlixPrizeData}.

\section{User profile Modelling Techniques}
\label{SEC:User_profiling}
To create a \textit{user profile}, we adopt the \textit{vector profile} representation, consisting of weighted attributes measuring the user's taste on each feature~\cite{DBLP:conf/cbmi/DeldjooCSQ17,DBLP:phd/hal/Kacem17}, because it is best suited for our evaluation in terms of similarity functions.
Formally, the user profiling methods we investigate build the user profile $\mathbf{p}_u$ as a vector whose attributes are the relevance weight of each feature $f$ for the target user $u$, denoted as~$h_{u,f}$.

We analyze 3 state-of-the-art methods from literature to model user profiles 
and we refer to them according to the first author of the corresponding publication, for simplicity and a 4\textsuperscript{th} method that applies the TF-IDF (term frequency--inverse document frequency) term weighting idea, which is widely used in CBF and, in general, in information retrieval~\cite{DBLP:reference/rsh/LopsGS11,DBLP:conf/hais/Sanchez-MorenoG18,DBLP:journals/kbs/WangLXFG18}.


\textbf{Zhang method.} Zhang et al.~\cite{DBLP:conf/aaai/ZhangWLXSY15} build the user profile based on item ratings or explicit feature ratings.
Let $\mathcal{U}$ and $\mathcal{I}$ denote the set of  users and items, respectively, and $\mathcal{F}$ the set of all features of the items. In case of binary ratings (like in our dataset), this method assigns relevance weight $h_{u,f}$ equal to 1 for each feature $f$ in $\mathcal{F}$ that applies to items with which the target user $u$ interacted with, 0 otherwise. 
The obvious limitation of this method is that it assigns only weights 0 or 1 to the features, without distinguishing their relevance for the user.

\textbf{Li method.} Li et al.~\cite{DBLP:conf/apweb/LiK04}, unlike Zhang et al., differentiate the relevance of features contained in an item by assigning scalar weights. Their method furthermore ignores items with low ratings by using a threshold value.
In case of binary ratings, the threshold rating $r_{\tau}$ 
is 0 and the relevance weight $h_{u,f}$ of each feature $f$ in $\mathcal{F}$ for the target user $u$ becomes the percentage of occurrences of $f$ in the items $u$ interacted with: 
$h_{u,f} = N_{u,f} / M_{u}$ 
, where $N_{u,f}$ is the number of items rated by user $u$ containing feature $f$ and 
$M_{u}$ is the total number of items rated by user $u$.


\textbf{Symeonidis method.} Symeonidis et al.~\cite{DBLP:conf/um/SymeonidisNM07} adopt an approach similar to TF-IDF to compute feature relevance weights, but define them in the vector space of 
user profiles.
The rationale of using TF-IDF is to increase the relevance of rare features contained in less user profiles. 
Symeonidis et al.~also use a fixed rating threshold to consider only the most relevant items. 
In case of binary ratings, the threshold rating $r_{\tau}$ is set to 0 and the relevance weight $h_{u,f}$ of each feature $f$ in $\mathcal{F}$ for the target user $u$ is computed as: $h_{u,i}= FF(u,f) \cdot IU\!F(f)$
, where $FF(u, f)$ is the feature frequency, i.e., the number of times feature $f$ occurs in movies rated by $u$, 
 and $IU\!F(f)$ is the inverse user frequency of feature $f$.  $IU\!F(f)= \log \frac{\left | \mathcal{U} \right |}{U\!F(f)}$
, where $U\!F(f)$ is the user frequency of $f\!$, i.e., the number of users whose rated movies contain feature $f$ at least once.


\textbf{TF-IDF method.} 
After having reviewed the 3 state-of-art methods described above, we decided to investigate another variant of TF-IDF as a user profiling method. The Symeonidis method above is similar to TF-IDF, but it is user-centric because it considers the vector space of user profiles. 
Instead, our proposed TF-IDF method is item-centric as it considers the vector space of items (movies).
First, we compute the IDF of each feature $f$ as: $ IDF(f) = \log \frac{\left | \mathcal{I} \right |}{n_f}$
, where $n_f$ denotes the number of items in $\mathcal{I}$ in which feature $f$ occurs at least once.
Then, for each user $u$, we compute the relevance weight $h_{u,f}$ of a feature $f$ as: $ h_{u,f} = TF(u,f) \cdot IDF(f)$
, where $TF(u,f)$ is equivalent to $FF(u, f)$ of the Symeonidis method (i.e., number of times feature $f$ occurs in items rated by user $u$). 
In contrast to the method by Symeonidis et al.,~$IDF(f)$ is computed in relation to all the existing items in which feature $f$ appears, not related to user profiles. 
As will be shown in Section~\ref{SEC:Evaluation:userProfiling}, our TF-IDF method yields better results than Symeonidis et al.'s.

\section{Data Acquisition}
\label{SEC:Experimental_Setup}


The dataset we use to evaluate user profiling methods has been collected through 
a web application we implemented, which can be navigated on a variety of stationary and mobile devices. 
It provides access to a large catalogue of more than 450K movies and any related content feature. This vast breadth of choice is possible thanks to the fact that we retrieve up-to-date information on-the-fly from TMDb\footnote{The Movie Database (TMDb): \url{www.themoviedb.org} 
} via APIs.
We developed the application with the idea of a completely free user experience, instead of making it like a survey application, so that users are not forced in any way during their selections.

To acquire the needed data, we asked users to select a set of ``favourites'', which included at least 5 movies, 2 genres, 3 actors, and 1 director. Users were, nevertheless, free to select more than these numbers of elements. 
We also asked users to provide some demographics information: age range, gender, and country of residence.
The collection of data was divided into two phases, the first one involved the \textit{volunteer} users, which are the ones invited to freely contribute (friends, family, acquaintances, and colleagues of the authors),
while the second phase involved users recruited by the \textit{crowdsourcing} platform MTurk,\footnote{Amazon Mechanical Turk (MTurk): \url{www.mturk.com}}
which have been paid between 20 and 50 US cents for their contribution. 
To assess the participants' reliability, we also asked them to complete a final consistency test 
which required to select again all (and only) the favourites they remember to have added (from a list of movies, genres, and actors of random popular elements).
A user's reliability is then estimated by means of the precision score 
computed on the re-selection of correct favourites. 

Finally, in order to explore 
a catalogue of existing features needed for user profiling evaluation, we retrieved The Movies Dataset~\cite{Banik2017TheMoviesDataset} containing the content of 45,3K movies scraped from TMDb. Then, we extended this dataset by scraping the content of missing movies that were added as favorites by users on our web application.


\textbf{Dataset characteristics.}
%
%
We have collected the preferences of 194 users, 180 (93\%) of whom have added the minimum number of required favourites.
Among all users, 81 (42\%) are volunteers and 113 (58\%) are paid ones. 
We consider users reliable if they are either volunteers that have completed the required favourites or crowdsourced users who scored at least 50\% of precision during the consistency test (see above). 
The reliable volunteers are 67 (83\% of all volunteers), while the crowdsourcing ones are 88 (78\% of all crowdsourcing), hence a total of 155 reliable users (80\% of all users).


Regarding users' gender, 115 users (59\%) are male, 66 are female (34\%), and 13 (7\%) did not specify gender. 53\% of the users are between 24 and 30 years old. We received registrations from users coming from 10 different countries, mainly from Italy (40\%), India (31\%), and United States (19\%). 

We collected a total 4,109 favourites (movies and content features) selected by participants, including 1,212 unique elements, i.e., favourites selected by at least one user. 
In the following experiments, we include only favourites of \textit{reliable} users, that are 3,341 (81\%), including 
1,737 favourite movies, 461 genres, 698 actors, 198 directors, 74 production companies, 92 production countries, 39 producers, 17 screenwriters, 21 release years, and 4 sound crew members.
The dataset is available on Kaggle\footnote{\url{https://www.kaggle.com/lucacostanzo/mints-dataset-for-recommender-systems}}. 




\section{Results and Discussion}
\label{SEC:Evaluation}
\subsection{Initial statistical analysis}


An initial statistical analysis highlights main differences between the set of all explicitly rated features and the set of all implicit features extracted from rated movies. 
 In Tables~\ref{table:results:commonGenres} and~\ref{table:results:commonActorsDirectors}, we present a comparison between the explicit and implicit sets of features, in percentage of common attributes (features), focusing on the $k$ most frequently selected attributes, respectively, for genre, actor, and director. These tables generally highlight a low overlap between the explicitly preferred features and the implicitly estimated ones (derived from favourite movies), in particular for actors and directors.
 The only exception is the genre attribute, which reveals a maximum overlap of 94.74\% when considering all 19 genres.
 These results generally confirm the previous findings in~\cite{Nasery2015PoliMovie:Systems} regarding existing gaps between explicitly selected features and implicitly estimated ones, with a different dataset containing more up-to-date movies and not limited to the most popular movies as used in~\cite{Nasery2015PoliMovie:Systems}.
\begin{table}[!hbpt]
\caption{\small{Common genres in the most selected $k$ attributes, either explicitly or implicitly
}}
\label{table:results:commonGenres}
\centering
\resizebox{.8\linewidth}{!}{%
\begin{tabular}{c|c|c}
\toprule
    $k$ & No.~common genres & \%~common genres\\
\midrule
       5 &       3 &      60.00\% \\
      10 &       8 &      80.00\% \\
      15 &      13 &      86.67\% \\
      All genres (19) &      18 &      94.74\% \\
\bottomrule    
\end{tabular}
}
\end{table}
%
%
%
%
%

%
\begin{table}[t!] 
\caption{\small{Common quota of either actors or directors between the most selected $k$ attributes, either explicitly or implicitly 
}}
\label{table:results:commonActorsDirectors}
\centering
\resizebox{.7\linewidth}{!}{%
\begin{tabular}{c|c|c}
\toprule
    $k$ & \% of common actors & \% of common directors\\
\midrule
      10 &       10.00\% &      30.00\% \\
      20 &       20.00\% &      20.00\% \\
      40 &      22.50\% &      27.50\% \\
      60 &      16.67\% &      28.33\% \\
\bottomrule    
\end{tabular}
}
\end{table}


We further provide a finer-grained analysis of the gap between explicit and implicit preferences of users according to their gender. 
In Tables~\ref{table:results:male:topFeatures} and~\ref{table:results:female:topFeatures}, we compare the 5 most frequently selected genres, actors, and directors, 
by \textit{male} and \textit{female} users, respectively. 
We notice a substantial difference between 
between male and female users with the exception of genre.

Investigating the results, it is surprising that in both  Tables~\ref{table:results:male:topFeatures}~and~\ref{table:results:female:topFeatures} Stan Lee is among the top  \textit{implicitly} preferred actors even if he barely acted as a main character in any movie. 
The most probable reason is that even though he has not been selected explicitly as favourite actor by study participants, he appeared in all 
Marvel movies (in small ``cameo roles''), so he is included in the implicit profiles.
Furthermore, it is surprising that the genre ``action'' is highly ranked by female users. This could be due to the fact that the genre tastes of young women might be changing nowadays, especially because many popular action movies, like the Marvel ones, are liked by many people (especially under 30, i.e., the largest age group in our dataset), irrespective of gender. Nonetheless, the other differences between male and female users suggest to embed gender information in a recommender system.

\begin{table}[!hbpt]
\caption{\small{Most selected 5 features, either explicitly ($ R_f^{exp}$) or implicitly ($ R_f^{imp} $), by male users;}}
\label{table:results:male:topFeatures}
\centering
\footnotesize
\resizebox{.8\linewidth}{!}{%
\begin{tabularx}{\linewidth}{l|r|X|c || X|c}
\toprule
    & Pos. & Explicit selection & $ R_f^{exp} $ & Implicit selection & $ R_f^{imp} $ \\
\midrule
  \multirow{5}*{Genres} & 1 &           Action &              51 &           Action &              86 \\
  & 2 &            Drama &              31 &        Adventure &              83 \\
  & 3 &        Adventure &              30 &            Drama &              80 \\
  & 4 &         Thriller &              28 &  Science Fiction &              76 \\
  & 5 &  Science Fiction &              28 &         Thriller &              74 \\
  
\midrule  
  
  \multirow{5}*{Actors} & 1 &  Robert Downey Jr. &              16 &  Samuel L. Jackson &              64 \\
  & 2 &        Johnny Depp &              15 &           Stan Lee &              56 \\
  & 3 &      Jason Statham &              10 &     Bradley Cooper &              51 \\
  & 4 &  Leonardo DiCaprio &              10 &       Paul Bettany &              47 \\
  & 5 &          Tom Hardy &               8 &         Vin Diesel &              47 \\
 
\midrule

  \multirow{5}*{Directors} & 1 &   Quentin Tarantino &              11 &         Hajar Mainl &              42 \\
  & 2 &    Steven Spielberg &               9 &      Chris Castaldi &              41 \\
  & 3 &           Joe Russo &               7 &        Mark Rossini &              41 \\
  & 4 &  M. Night Shyamalan &               6 &      Lori Grabowski &              41 \\
  & 5 &   Christopher Nolan &               6 &          Eli Sasich &              41 \\
  
\bottomrule    
\end{tabularx}
}
\end{table}

\begin{table}[!hbpt]
\caption{\small{Most selected 5 features, either explicitly ($ R_f^{exp}$) or implicitly ($ R_f^{imp} $), by female users;}}.
\label{table:results:female:topFeatures}
\centering
\footnotesize
\resizebox{.8\linewidth}{!}{%
\begin{tabularx}{\linewidth}{l|r|X|c || X|c}
\toprule
    & Pos. & Explicit selection & $ R_f^{exp} $ & Implicit selection & $ R_f^{imp} $ \\
\midrule
  \multirow{5}*{Genres} & 1 &            Drama &              26 &            Drama &              52 \\
  & 2 &           Action &              22 &        Adventure &              48 \\
  & 3 &        Adventure &              14 &           Action &              47 \\
  & 4 &           Comedy &              14 &          Fantasy &              45 \\
  & 5 &         Thriller &              13 &  Science Fiction &              43 \\
  
\midrule  
  
   \multirow{5}*{Actors} &    1 &     Robert Downey Jr. &              12 &              Stan Lee &              27 \\
  & 2 &     Leonardo DiCaprio &               7 &     Samuel L. Jackson &              26 \\
  & 3 &     Jennifer Lawrence &               5 &        Bradley Cooper &              23 \\
  & 4 &       Chris Hemsworth &               5 &        Djimon Hounsou &              21 \\
  & 5 &          Bruce Willis &               4 &          James McAvoy &              21 \\
 
\midrule

  \multirow{5}*{Directors} & 1 &           Joe Russo &               4 &       Anthony Russo &              16 \\
  & 2 &   Christopher Nolan &               4 &           Joe Russo &              16 \\
  & 3 &    Steven Spielberg &               4 &        Bryan Singer &              15 \\
  & 4 &     Martin Scorsese &               2 &         Hajar Mainl &              14 \\
  & 5 &        Ridley Scott &               2 &      Chris Castaldi &              14 \\  
  
\bottomrule    
\end{tabularx}
}
\end{table}

\subsection{Evaluation of user profiling methods}\label{SEC:Evaluation:userProfiling}
We study the user profiling step in-depth by investigating the 4 user profiling methods described in Section~\ref{SEC:User_profiling}. Our aim is to analyze the similarity (i.e., the overlap) between the implicitly modelled user profiles and the real explicit tastes of users.
For each target user $u$, we built his or her \textit{explicit profile} $\mathbf{p}'_u$ as vector composed of relevance weights equal to 1, for all the features explicitly rated by $u$, and weight 0 for the ones not rated. Then we computed the pairwise similarity between the explicit user profiles and  \textit{implicit profiles} $\mathbf{p}_u$ produced by each method, using cosine similarity and Jaccard 
similarity. The highest is this similarity, the most accurate is the implicit user profile modelled.

The average pairwise similarity $sim(\mathbf{p}_u, \mathbf{p}'_u)$ between implicit user profile $\mathbf{p}_u$ and explicit one $\mathbf{p}'_u$ is shown in Table~\ref{table:results:averagePairwiseSimilarity}.
As revealed in the table and already anticipated in Section~\ref{SEC:User_profiling}, the TF-IDF method yields better results than Symeonidis even if they are intrinsically similar, hence the item-centric TF-IDF approach outperforms the user profle-based one. 
In general, the average pairwise similarities are remarkably low, even for the best investigated method, i.e.,~Li.
The overlap between explicit and implicit profiles increases if we consider only genres; the reason is that the catalogue of all possible genres in the dataset is rather limited (19) compared to actors (567K) and directors (58K). The Jaccard measure yields lower similarities because it can be applied only to vectors composed of binary attributes 
while our tested profiling methods compute scalar weights (except for Zhang); hence we had to cut-off some feature weights by considering only the k most relevant features in the implicit profile of each user considered, in which k is the number of explicit features rated by that user.

The presented results underline the low effectiveness of the investigated user profiling methods to model real user tastes. This finding gives rise to 
the need of further research on this important user profiling step when devising recommender systems. If user profiles are not properly modelled before applying any RS technique, the accuracy of the final recommendations will likely be affected and lowered by an inaccurate representation of the user's tastes.


\begin{table}[!t]
\caption{\small{Average pairwise similarity between \textit{explicit} and \textit{implicit} user profiles, for all the methods.}}
\label{table:results:averagePairwiseSimilarity}
\centering
\resizebox{.8\linewidth}{!}{%
\begin{tabular}{l|l|r|r|r|r}
\toprule
Similarity & Feature & Zhang & Li & Symeonidis & TF-IDF \\
\midrule
\multirow{3}*{Cosine} & Genre      &  48.52\% &  58.07\% &       42.00\% &  53.08\% \\
& Actor      &   7.03\% &   9.13\% &        6.50\% &   7.24\% \\
& Director  &  15.17\% &  17.24\% &       15.32\% &  16.14\% \\

\midrule

\multirow{3}*{Jaccard} & Genre      &  27.49\% &  36.19\% &       18.54\% &  33.36\% \\
& Actor     &   0.97\% &   5.73\% &        2.87\% &   4.64\% \\
& Director  &   5.22\% &  10.24\% &        6.30\% &   8.17\% \\
\bottomrule
\end{tabular}
}
\end{table}





\section{Conclusion and Future Works}\label{SEC:CONCLUSION}
In this paper, we analyzed the user profiling modelling by studying the differences between explicit user preferences and implicit user profiles. We evaluated different user profiling methods and showed that even the best profiling method that we tested provided low pairwise similarities between explicit and implicit profiles. This finding can be explained by the fact that when a user rates a movie, he is implicitly rating only some characteristics of the item that impacted on her (but not all). Also, it could happen that a user may select a movie but she only loved some part of it (e.g., very good director but bad actors), and this can result in the introduction of some noise in the learning process. Overall, our study encourages a more in-depth on ways we can obtain reliable feedbacks on features and study the optimization of the user profile modelling step in RS, which will eventually allow to produce more accurate recommendations. Furthermore, we publicly provide the dataset that we collected and used for evaluation, which includes ratings on movies and on corresponding content features. 

In the future, we plan to investigate the generalizability of findings in this work on other domains where the exist a wide variety of item content features and personalization on these features is paramount, in domains including but not limited to fashion~\cite{he2016ups}, music domain~\cite{schedl2018current}, tourism~\cite{adamczak2019session,Knees_etal:RSC:2019} and so forth.

\bibliographystyle{ACM-Reference-Format}
\bibliography{recsys2019} 

\end{document}